\documentclass[letterpaper,10pt,conference]{ieeeconf}
\IEEEoverridecommandlockouts
\usepackage[utf8]{inputenc}
\usepackage{cite}
\usepackage{graphics}
\usepackage{graphicx}

\usepackage[caption=false]{subfig}

\usepackage{comment}
\usepackage{amsmath,amsfonts}
\usepackage{amssymb}
\usepackage{cite}
\usepackage{algorithmic}
\usepackage{algorithm}
\usepackage[T1]{fontenc}
\usepackage{textcomp}
\usepackage{stmaryrd}
\usepackage{amsthm}
\usepackage{siunitx}
\usepackage{booktabs}
\usepackage{xcolor}
\newtheorem{definition}{Definition}  
\newtheorem{problem}{Problem}
\newtheorem{theorem}{\textbf{Theorem}}
\newtheorem{remark}{Remark}

\newtheoremstyle{mystyle}
  {}
  {}
  {}
  {}
  {\bfseries}
  {.}
  { }
  {}

\theoremstyle{mystyle}

\usepackage{titlesec}
\titlespacing{\section}{0pt}{0.2ex}{0.2ex}
\titlespacing{\subsection}{0pt}{0.1ex}{0.1ex}
\titlespacing{\subsubsection}{0pt}{0.1ex}{0.1ex}

\newcommand{\zono}[1]{\left\langle#1\right\rangle}

\newcommand{\enc}[1]{\llbracket #1 \rrbracket}

\title{\LARGE \bf Privacy Guarantees for Cloud-based State Estimation using Partially Homomorphic Encryption
  \thanks{$^1$The authors are with Computer and Systems Department, Ain Shams University. \texttt{\{sawsan.emad, watheq.elkharashi\}@eng.asu.edu.eg}. $^2$The author is with Jacobs University, Bremen. \texttt{a.alanwar@jacobs-university.de}. $^3$The author is with Halmstad University. \texttt{yousra.alkabani@hh.se}.
    $^4$The authors are with KTH Royal Institute of Technology. \texttt{\{hsan, kallej\}@kth.se}. 
  }
  \thanks{This work was supported by the Swedish Research Council, the Knut and Alice Wallenberg Foundation, the Democritus project on Decision-making in Critical Societal Infrastructures by Digital Futures, and the European Unions Horizon
   2020 Research and Innovation program under the CONCORDIA cyber security project (GA No. 830927).
  }
}

\author{Sawsan Emad$^1$, Amr~Alanwar$^2$, Yousra~Alkabani$^{3}$, M. Watheq El-Kharashi$^1$,\\ Henrik~Sandberg$^4$, and Karl~Henrik~Johansson$^4$}

\begin{document}

\maketitle


\begin{abstract}

The privacy aspect of state estimation algorithms has been drawing high research attention due to the necessity for a trustworthy private environment in cyber-physical systems. These systems usually engage cloud-computing platforms to aggregate essential information from spatially distributed nodes and produce desired estimates. The exchange of sensitive data among semi-honest parties raises privacy concerns, especially when there are coalitions between parties. We propose two privacy-preserving protocols using Kalman filter and partially homomorphic encryption of the measurements and estimates while exposing the covariances and other model parameters. We prove that the proposed protocols achieve satisfying computational privacy guarantees against various coalitions based on formal cryptographic definitions of indistinguishability. We evaluate the proposed protocols to demonstrate their efficiency using data from a real testbed. 
 

\end{abstract}

\label{abstract}
\section{Introduction}\label{sec_introduction}

Cyber-physical systems (CPSs) have emerged as the new paradigm for the modern global technology industry, representing a highly interactive generation of intelligent systems with tight integration between computer resources and physical processes \cite{CPSreview}. 
Some states of these systems aren't directly perceptible by sensors; sensors may be unable to sense data from the area of interest or can only sense physical variables relevant to the variables of interest, or measurements may be inaccurate or subject to noises \cite{CPScomprehensive}.
To maintain robustness against measurement noise and modeling uncertainty, optimal state estimation algorithms that implement multisensor data fusion \cite{multisensorfusion} are employed to find the best estimates for hidden states with minimal estimation error \cite{estimation}.

Typically, the estimator (aggregator) aggregates essential information from spatially distributed sensors, applies an estimation algorithm to produce the required estimates, and then sends them to the interested party who initiated the inquiry. Thus, estimators are usually outsourced to cloud-computing platforms like in \cite{sets_est_privacy,hierEst_privacy_noKF} and can also be centralized or distributed among multiple nodes \cite{est_distr,est_distr2}.
Kalman filters \cite{first_kalman} are widely-used optimal estimation algorithms that can fuse measurements and estimates \cite{KF_msensors,1st_msensor_kalman} within centralized or distributed implementations \cite{KF,KFdistr} and provide accurate and precise estimates of hidden states considering process and measurement uncertainties.

Because cloud-based estimations use open computation and communication architectures, they might suffer from adversarial physical faults or cyber-attacks. 
Therefore, researchers proposed several approaches to perform computations on sensitive data while keeping the data confidential from untrustworthy parties, such as differential privacy \cite{differentialKF,differentialKF_error}, obfuscation \cite{obfuscate1,obfuscate2}, algebraic transformation \cite{KF_transf_privacy,ctrl_transf_privacy} and homomorphic encryption \cite{mparty_privacy_opt_coal,mparty_privacy_opt}. Paillier encryption, which is partially homomorphic encryption (PHE), was employed with several estimation algorithms to preserve data privacy as in \cite{sets_est_privacy,privateEst_noKF,hierEst_privacy_noKF}. 
Kalman filters can operate in an encrypted domain while retaining their natural effectiveness. %
A secure state estimation using Kalman filter with the adoption of a hybrid homomorphic encryption scheme 
was proposed in \cite{centKF_hybridEnc}. Authors in \cite{distrKF_PHE} presented a multi-party dynamic state estimation using the Kalman filter and PHE, while \cite{secure_distrKF} introduced a secure distributed Kalman filter using PHE
. However, no work to date has provided a computational privacy investigation for estimation algorithms that use Kalman filters along with PHE, considering that other problems have undergone similar computational privacy analysis, such as set-based estimation in \cite{sets_est_privacy} and quadratic optimization in \cite{mparty_privacy_opt}.

We focus on the privacy of multi-party cloud-based state estimation of a linear discrete time-invariant (LTI) system where the involved parties communicate over end-to-end encrypted networks.  
We consider semi-honest parties that follow protocols properly but keep a record of all their intermediate computations and may collude with other parties to reveal private information of non-colluding ones. 
In short, we make the following contributions: 
\begin{itemize}
    \item We propose two privacy-preserving estimation protocols using Kalman filters and Paillier cryptosystem by encrypting the measurements and estimates while revealing their covariances and model parameters. 
    \item We provide computational privacy guarantees for the proposed protocols against various coalitions of semi-honest parties using formal cryptographic definitions of computational indistinguishability.
\end{itemize}

The remainder of this paper is organized as follows. 
We demonstrate two problem setups in Section \ref{sec_problem_setup} and follow them with privacy definitions and preliminaries in Section \ref{sec_privacy_goals}. 
We propose two privacy-preserving protocols and summarize their privacy guarantees in Sections \ref{sec_proposed_solution_1} and \ref{sec_proposed_solution_2}. Then, we discuss the protocols' privacy guarantees in Section \ref{sec_privacy_discussion}. 
Finally, we evaluate the proposed protocols in Section \ref{sec_evaluation} and conclude this paper in Section \ref{sec_conclusions}.
\section{Problem Setup} \label{sec_problem_setup}


We consider two common problem setups similar to \cite{sets_est_privacy}. The first setup is in Fig. \ref{fig_problem1_setup}, and it involves:

\begin{figure}
    \centering
     \includegraphics[width=\linewidth]{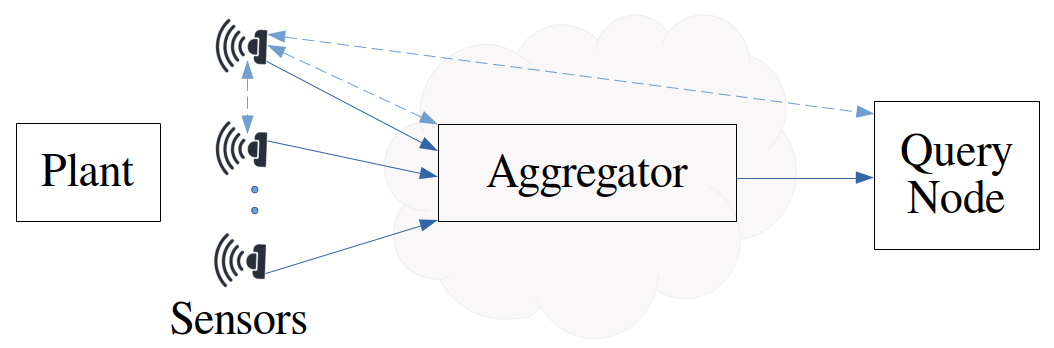}
      \caption{Problem \ref{prob1_stmnt} setup where the bold arrows represent information communication and the dashed arrows represent coalitions.}
      \label{fig_problem1_setup}
  \vspace{-4mm}
\end{figure}

\begin{itemize}
    \item  
        \textbf{Plant $T$:} A passive entity whose states need to be estimated. We consider the state estimation of a plant modeled as a 
        linear discrete time-invariant (LTI) dynamic system whose state-space model of the form:
        \begin{align}
            \mathbf{x}_{k+1} &= \mathbf{F} \mathbf{x}_k + \mathbf{n}_k, \\ 
            \mathbf{y}_{i,k} &= \mathbf{H}_i \mathbf{x}_k + \mathbf{v}_{i,k}, 
            \label{eq_sys}
        \end{align}
        
        where $ \mathbf{x}_k \in \mathbb{R}^n $ is the system state at time step $ k \in \mathbb{N} $,
        $ \mathbf{y}_{i,k} \in \mathbb{R}^p $ the measurements of sensor $ i \in {1, \dots , I}$, 
        $\mathbf{F} \in \mathbb{R}^{n \times n} $ the process matrix, $ \mathbf{H} \in \mathbb{R}^{p \times n} $ the measurement matrix,
        $ \mathbf{n}_k \in \mathbb{R}^n $ the modeling noise and 
        $ \mathbf{v}_{i,k} \in \mathbb{R}^p$ the measurement noise and both 
        are independent zero-mean
        Gaussian white noises with covariances $ \mathbb{\mathbf{Q}}_k \in \mathbb{R}^{n \times n} $ and $\mathbf{R}_{i,k} \in
        \mathbb{R}^{p \times p} $ respectively. 
        
    \item  
        \textbf{Sensor $S_i$:} An entity with index i that provides measurements containing sensitive information that should not be revealed to other parties.
    
    \item  
        \textbf{Aggregator $A$ (or Cloud):} An untrusted party has reasonable computational power that is needed to implement the estimation protocols. 
        It collects encrypted data synchronously from other parties and operates in an encrypted domain to provide the query node with encrypted estimates of the plant $T$ states.
        
    \item  
        \textbf{Query Node $Q$:}
        An untrusted party inquires about private states of plant $T$, which no other party has the right to know. Besides, it owns the encryption keys and shares the public key $pk$ with others while keeping the private key $sk$ hidden. The query node can be any entity other than the aggregator $A$, including the plant $T$.
    
\end{itemize}

Briefly, we seek to solve the following first problem:
\begin{problem}
\label{prob1_stmnt}
How to ensure privacy is preserved while estimating the plant $T$ states by a remote aggregator $A$ using measurements of spatially distributed sensors? It is required to ensure that measurements are private to the sensor nodes $S_1, \dots , S_I$ and the estimated states are private to the query node $Q$, and to guarantee computational security during the estimation process as well.
\end{problem}
The second problem setup is in Fig. \ref{fig_prob2_setup},
which includes the following entities in addition to the predefined entities:   

\begin{figure}
    \centering
    \includegraphics[width=\linewidth]{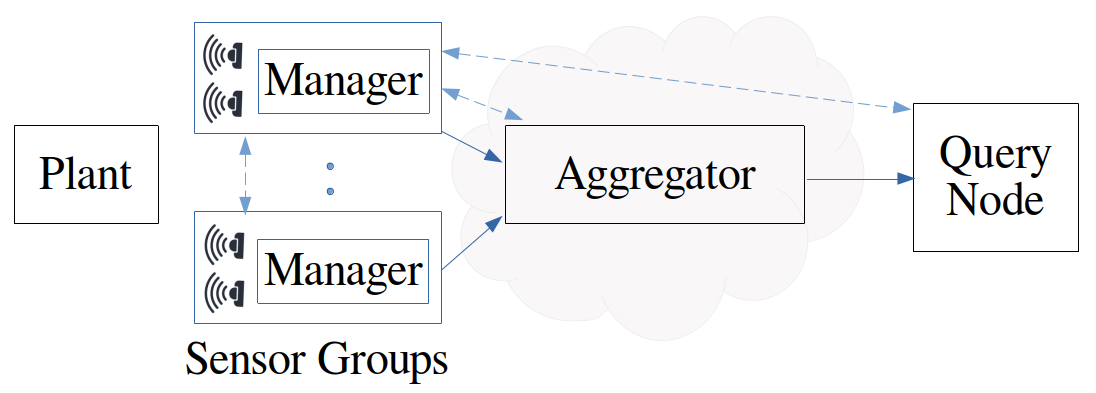}
    \caption{Problem \ref{prob2_stmnt} setup where the bold arrows represent information communication and the dashed arrows represent coalitions.}
	\label{fig_prob2_setup}
	\vspace{-4mm}
\end{figure}

\begin{itemize}
\item \textbf{Manger ${M}_j$:} 
An entity with index $j$ produces local estimates of plant $T$ states using synchronously collected measurements from sensors within the group and handles communication with entities outside the group.
\item \textbf{Sensor Group $G_j$:} 
An entity with index $j$ includes one manager $M_j$ and $I_{g_j}$ sensors, all owned by one organization. All group members trust each other while each group aims to keep its measurements and estimates private from other groups/parties. 
\end{itemize}

The problem statement of the second setup is:
\begin{problem}
How to ensure privacy is preserved in a multi-party cloud-based estimation? It is required to ensure that measurements and local estimates are private to sensor groups ${G}_1, \dots , {G}_J$ and that global estimates are 
private to query node $Q$ and sensor groups, 
and to guarantee computational security during the estimation process as well.
\label{prob2_stmnt}
\end{problem}
The global estimate is computed by the aggregator node using local estimates provided by the sensor groups.
We should note that 
Problem \ref{prob1_stmnt} is not a special case of Problem \ref{prob2_stmnt} because of the existence of group managers \cite{sets_est_privacy}. 
In sections \ref{sec_proposed_solution_1} and \ref{sec_proposed_solution_2}, we present two protocols to provide private solutions for the above problems, and the purpose of this paper is to ensure that the proposed protocols preserve privacy against the following coalitions \cite{sets_est_privacy}: 

\begin{definition} 
    [\textbf{Sensor coalition}] \label{def_sensor_caol}
    A number of sensors/sensor groups $t$ collude together by exchanging their private measurements to infer private information of the query node and non-colluding sensors/sensor groups. 
\end{definition}
\begin{definition} 
    [\textbf{Cloud coalition}] \label{def_cloud_caol}
    Aggregator ${A}$ colludes with up to $t$ sensors/sensor groups by exchanging private information and intermediate results to infer private information of the query node and non-colluding sensors/sensor groups.
\end{definition}
\begin{definition} 
    [\textbf{Query coalition}] \label{def_query_caol}
  Query node ${Q}$ colludes with up to $t$ sensors/sensor groups by exchanging their private information, private cryptographic keys and the decrypted results to infer private information of non-colluding sensors/sensor groups.
\end{definition}
These coalitions are visualized by dashed lines in Fig.~\ref{fig_problem1_setup} and Fig.~\ref{fig_prob2_setup}. 
  We assume that only one type of the above coalitions can occur at a time, and there is always at least one sensor/sensor group that does not participate in the coalition. 

In the next section, we outline the privacy goals we rely on to demonstrate that our protocols preserve privacy against the above coalitions.

\section{Privacy Goals and Preliminaries} \label{sec_privacy_goals}
We start by defining our privacy goals.
\subsection{Privacy Goals}
To preserve data privacy, our protocols must guarantee computational security against predefined coalitions. 
In other words, privacy is preserved if all that a coalition of parties can obtain from 
keeping records of the intermediate computations can essentially be obtained from these parties' inputs and outputs only \cite[p.~620]{conf_cryptoref2}. Moreover, the coalition's inputs and outputs and recorded information cannot be exploited to infer further private information \cite{sets_est_privacy}. These privacy goals are based on the 
following definitions. 

    Let $\{0,1\}^\star$ be a sequence of bits with indefinite length. Then, an ensemble $X =\{X_n\}_{n \in \mathbb{N}}$ is a sequence of random variables $X_n$ that ranges over strings of bits with a length polynomial in $n$.
    
    \begin{definition}\label{def_comp_indist}(\!\!\cite[p.105]{conf_cryptoref})
        (\textbf{Computationally Indistinguishable})
        The ensembles $X = \{X_n\}_{n \in \mathbb{N}}$ and $Y = \{Y_n\}_{n \in \mathbb{N}}$ are computationally indistinguishable, denoted as $X\stackrel{c}{\equiv}Y$, if for every probabilistic polynomial-time algorithm $D$, each positive polynomial $p(.)$ with all sufficiently large $n$'s, it follows 
        \begin{align}
            | \text{Pr}[D(X_n)=1] - \text{Pr}[D(Y_n)=1] | < \frac{1}{p(n)}.
        \end{align}
    \end{definition}
    
    
    \begin{definition}(\!\!\cite[p.620]{conf_cryptoref2}) \label{def_exec_view}
        (\textbf{Execution View})
        Let $f(\bar{x})=(f_1(\bar{x}),\dots,f_n(\bar{x}))$ be a deterministic polynomial-time function and $\Pi$ a multi-party protocol that computes $f(\bar{x})$ with the input $\bar{x} = ( x_1,\dots,x_n )$. The view of the $i^{th}$ party during an execution of $\Pi$ using $\bar{x}$, is defined as 
        \begin{align}
            V^{\Pi}_i(\bar{x}) =  (x_i,coins,M_i),
        \end{align}
        where $coins$ are the outcome of the party’s internal coin toss, and $M_i$ is the set of messages it has received. For coalition $I = {i_1,\dots,i_t } \subseteq \{1,\dots,n \}$ of parties, the coalition view $V^{\Pi}_I(\bar{x})$ during an execution of $\Pi$ is defined as \cite[p.696]{conf_cryptoref2}
        \begin{align} 
        V^{\Pi}_I(\bar{x}) = \big(V^{\Pi}_{i_1}(\bar{x}),\dots,V^{\Pi}_{i_t}(\bar{x})\big).
        \end{align}
    \end{definition}
    
    
    \begin{definition}(\!\!\!\!\cite{sets_est_privacy}) \label{def_semihonest}
        \textbf{(Multi-Party Privacy w.r.t. Semi-Honest Behavior})
        Considering the coalition of parties $I = \{i_1,\dots,i_t \} \subseteq \{1,\dots,n \}$, we have $\bar{x}_I = (x_{i_1},\dots,x_{i_t})$ and $f_I(\bar{x}) = \big(f_{i_1}(\bar{x}),\dots,f_{i_t}(\bar{x})\big)$,
        where $f(\bar{x})$ is a deterministic polynomial-time function. We say that the multi-party protocol $\Pi$ computes $f(\bar{x})$ privately if  
        \begin{itemize}
            \item there exists a probabilistic polynomial time algorithm, denoted by simulator $S$, such that for every $I \subseteq \{1,\dots,n \}$\cite[p.696]{conf_cryptoref2}:
            \begin{align}
              S\big( \bar{x}_I,f_I(\bar{x}) \big)  \stackrel{c}{\equiv} V^{\Pi}_I(\bar{x}),
            \end{align}
            \item the inputs and outputs of the coalition cannot be exploited to infer further private information. 
        \end{itemize}
    \end{definition}
    
    Thus, our proofs in Appendix \ref{app1} and \ref{app2} will consist of proving these two points of Definition \ref{def_semihonest}.
    



\subsection{Paillier Homomorphic Cryptosystem} \label{sec_paillier_crypto}
The homomorphic cryptosystem is a cryptographic primitive that allows computation over encrypted data. 
Proposed protocols use Paillier additive homomorphic cryptosystems~\cite{phe_paillier}, which is a probabilistic public-key cryptography scheme
 that provides two basic operations, namely (i) addition and subtraction of two encrypted values denoted by $\oplus$ and $\ominus$, respectively, and (ii) multiplication of an encrypted value by a plaintext value denoted by $\otimes$. That is, if we denote the encryption of $a$ using the public key $\textit{pk}$ by $\enc{a}$, then the Paillier cryptosystem supports
\begin{align}
&\textsc{Decrypt}_{\text{sk}}(\enc{a} \oplus \enc{b}) = a + b, \label{eq:pailadd}\\
& \textsc{Decrypt}_{\text{sk}}(a \otimes \enc{b}) = a \times b, \label{eq:pailmul}
\end{align}
where $\textit{sk}$ is the private key associated with public key $\textit{pk}$. 
We will omit the symbol $\otimes$ when the type of multiplication can be inferred from the context. 
The security guarantees of the Paillier cryptosystem rely on the standard cryptographic assumption named decisional composite residuosity assumption (DCRA)~\cite{phe_paillier}. We use the floats encoding mechanism presented in \cite{encoding_localization} that represents float numbers by a positive exponent and a mantissa, which are both integers and thus can be used with the aforementioned cryptosystems \cite{conf_crypt}.

In the next two sections, we present our privacy-preserving state estimation protocols using synchronously collected measurements. 
We consider that all parties communicate over end-to-end encrypted channels. In our protocols, we propose using the Paillier cryptosystem to implement further encryption only for measurements and estimates, without covariances and model parameters due to the nature of the PHE. We don't use the fully homomorphic encryption FHE as PHE is more efficient and good enough since knowing only covariances doesn't reveal measurements or estimates. 
For simplicity, we will denote only paillier encryption by $\enc{ }$.
\section{Private estimation among distributed sensors} \label{sec_proposed_solution_1}


We propose a protocol that solves Problem \ref{prob1_stmnt} to estimate the states of a system within the first pre-stated setup in Section \ref{sec_problem_setup} while preserving information privacy.

Initially, the query node generates the Paillier public key $pk$ and the private key $sk$ and shares the public key with other parties, then sends the initial estimates to the aggregator after encrypting its state vector $\enc{{\mathbf{x}}_{q,0}}$ while revealing its covariance matrix $\mathbf{P}_{q,0}$. At every time step $k$, each sensor $i$ also encrypts its measurement vector $\enc{\mathbf{y}_{i,k}}$ and reveals its covariance matrix $\mathbf{R}_{i,k}$, and then sends them to the aggregator. The aggregator, in turn, aggregates all received measurements and applies the estimation algorithm steps \cite[p.~190]{book_KF}. 
During the time update, the Kalman filter produces a predicted estimate for the system states in the current time step. During the measurement update, the predicted estimate is updated by processing all measurements in parallel. 
All measurement vectors are combined to form a new measurement vector $\enc{\mathcal{Y}_{k}} \in\mathbb{R}^{ pI }$ where $I$ is the total number of sensors. Similarly, $\mathcal{H} \in\mathbb{R}^{ pI \times n}$ is the new observation matrix, and assuming that measurement noises are uncorrelated, the covariance of $\enc{\mathcal{Y}_{k}}$ is the diagonal matrix ${\mathcal{R}_{k}} \in\mathbb{R}^{pI \times pI}$. A new gain matrix ${\mathcal{K}_{k}} \in\mathbb{R}^{n \times pI}$ is computed and then used to find the optimal estimate that the query node inquires. 
As the query node has the private key $sk$, it decrypts the received estimates and retrieves the desired state estimate ${\mathbf{x}}_{q,k}$ for each time step. All the computational steps are detailed in Protocol \ref{protocol_1}.
\begin{algorithm}[t]
\floatname{algorithm}{Protocol}
\caption{Private Estimation Among Distributed Sensors}
\label{protocol_1}
\begin{algorithmic}
    \STATE The query node ${Q}$ encrypts the initial state $\enc{{\mathbf{x}}_{q,0}}$ and sends it with the initial error covariance $\mathbf{P}_{q,0}$ (not encrypted) to the aggregator node ${A}$ to have $\enc{\hat{\mathbf{x}}_{a,0}}=\enc{{\mathbf{x}}_{q,0}}$
    and ${\mathbf{P}}_{a,0}={\mathbf{P}}_{q,0}$, 
    \STATE At every time step $k$, every sensor node shares its encrypted measurements $\enc{\mathbf{y}_{i,k}}$ and the measurement noise covariance matrix $\mathbf{R}_{i,k}$ (not encrypted)
    with the aggregator.\\
    In order to find the required estimates, the aggregator uses Kalman filter estimation algorithm as following:
    
    \STATE \textbf{Step 1}: Time update at the aggregator:
    \begin{align} \label{eqn_enc_kalman1}
        & \enc{\hat{\mathbf{x}}_{a,k}^{-}} = \mathbf{F} \enc{\hat{\mathbf{x}}_{a,k-1}}, 
    \\
     \label{eqn_enc_kalman2}
        & \mathbf{P}^{-}_{a,k} = \mathbf{F} \mathbf{P}_{a,k-1} \mathbf{F}^{T} + \mathbf{Q}_{k}.
    \end{align}

    \STATE \textbf{Step 2}: Measurement update at the aggregator:
    \begin{align} \label{eqn_enc_kalman3}
        & \mathbf{P}_{a,k} = \left( (\mathbf{P}^{-}_{a,k})^{-1} + \mathcal{H}_{k}^{T}  \mathcal{R}_{k}^{\mathcal{y}} \mathcal{H}_{k}  \right)^{-1},
    \\
    \label{eqn_enc_kalman4}
        & \mathcal{K}_{k} = \mathbf{P}_{a,k} \mathcal{H}_{k}^{T}  \mathcal{R}_{k}^{\mathcal{y}},
        \\
    \label{eqn_enc_kalman5}
        & \enc{\hat{\mathbf{x}}_{a,k}} = \enc{\hat{\mathbf{x}}_{a,k}^{-}} \oplus \mathcal{K}_{k} \left( \enc{\mathcal{Y}_{k}} \ominus \mathcal{H}_{k} \enc{\hat{\mathbf{x}}_{a,k}^{-}}\right).
    \end{align}
    \begin{align}
    \nonumber
    & where \; 
    \\ \nonumber
    & \enc{\mathcal{Y}_{k}} = \left[ \enc{\mathbf{y}_{1,k}^{T}}, \dots, \enc{\mathbf{y}_{I,k}^{T}}\right]^{T}, 
    \\ \nonumber
    & \mathcal{H}_{k} = [ \mathbf{H}_{1}^{T}, \dots ,\mathbf{H}_{I}^{T}]^{T}, 
    \\ \nonumber
    & \mathcal{R}_{k} = diag \: \{ \mathbf{R}_{1,k}, \dots ,\mathbf{R}_{I,k} \}, 
    \\ \nonumber
    & \mathcal{y} \: \text{represents the Moore–Penrose pseudoinverse},
    \\ \nonumber
    & \mathcal{K}_{k} = [ \mathbf{K}_{1,k},  \dots ,\mathbf{K}_{I,k}] 
    \: \: and \: \:  \nonumber
    \mathbf{K}_{i,k} = \mathbf{P}_{a,k} \mathbf{H}_{i}^{T}  \mathbf{R}_{i,k}^{-1}.
    \end{align}

    \STATE \textbf{Step 3}: The aggregator sends the  outputs $\enc{\hat{\mathbf{x}}_{a,k}}$ and $\mathbf{P}_{a,k}$ to the query node to have $\enc{\hat{\mathbf{x}}_{q,k}}=\enc{\hat{\mathbf{x}}_{a,k}}$
    and ${\mathbf{P}}_{q,k}={\mathbf{P}}_{a,k}$. In turn, query node decrypts $\enc{\hat{\mathbf{x}}_{q,k}}$ and gets the desired estimate $\hat{\mathbf{x}}_{q,k} $.

\end{algorithmic}
%
\end{algorithm}
\setlength{\textfloatsep}{8pt}
In the following theorem, we summarize the privacy guarantees of the protocol against coalitions predefined in Definitions \ref{def_sensor_caol}, \ref{def_cloud_caol}, and \ref{def_query_caol}: 
\begin{theorem}
\label{th_dist_setup}
Protocol \ref{protocol_1} solves Problem \ref{prob1_stmnt} while preserving computational privacy against
\begin{itemize}
    \item Sensor coalitions,
    \item Cloud coalitions,
    \item Query coalitions if $pm_r>n$, where $p$ is the measurement size, $m_r$ the number of non colluding sensors and $n$ the state size.
\end{itemize}
\end{theorem}
Proof of Theorem \ref{th_dist_setup} is detailed in Appendix \ref{app1}.

\section{Private estimation among sensor groups} \label{sec_proposed_solution_2}

In this section, we present a protocol to solve Problem \ref{prob2_stmnt} using a diffusion Kalman filter algorithm. Unlike Protocol \ref{protocol_1}, where both estimates and measurements are encrypted, now only estimates are homomorphically encrypted. There is no need to encrypt measurements within the same group as 
each sensor trusts all other sensors within the same group.
Initially, the query node shares the initial estimates only with sensor groups, while it shares the Paillier public key with all parties after generating a pair of keys.

At every time step $k$, each sensor $i$ within group $j$ participates with its measurements $\mathbf{y}_{i,k}$. All measurements within-group are collected and used along with the previously estimated state $\hat{\mathbf{x}}_{q,k-1}$ to compute a new prior estimate $\hat{\mathbf{x}}_{{g_{j}},k}^{-}$. 
Then, the owner of each sensor group encrypts its prior estimate $\enc{\hat{\mathbf{x}}_{{g_{j}},k}^{-}}$ homomorphically and sends it to the aggregator. The aggregator, in turn, performs the diffusion update and computes a weighted average of all received encrypted estimates based on their uncertainties assuming that they are independent from each other which implies zero cross-covariances \cite{book_distKF}. Next, the aggregator calculates the time update to find and submit the 
optimal estimate for the current time step 
$\enc{\hat{\mathbf{x}}_{a,k}}$. Finally, the query node decrypts the received result to find the desired estimated state $\hat{\mathbf{x}}_{q,k}$ and then sends it to all sensor groups for use in the next iteration. 

    Unlike Protocol \ref{protocol_1}, which performs both Kalman filter steps (measurement update and time update) at the aggregator in the encrypted domain, Protocol \ref{protocol_2} performs the measurement update at the sensor group level in the plaintext domain and implements a diffusion update step
     along with the time update step
     at the aggregator in the encrypted domain. For the diffusion update, there are several methods in the literature \cite{KFdistr}, and we chose the weighted average method since it suits homomorphic computations.
 \begin{algorithm}[t]
\floatname{algorithm}{Protocol}
\caption{Private Estimation among Sensor Groups}
\label{protocol_2}
\begin{algorithmic}
\STATE Query node ${Q}$ sends the initial state to all sensor groups. At every time instant $k$, following steps are executed: \\
 \STATE \textbf{Step 1}: Measurement update at each sensor group $j$:
\begin{align}
      & \mathbf{P}^{-}_{g_{j},k} = \left( (\mathbf{P}_{q,k-1})^{-1} + \sum\limits_{i=1}^{I_j}{\mathbf{H}_{i}^{T} \mathbf{R}_{i,k}^{-1} \mathbf{H}_{i} } \right)^{-1},
      \label{eqn_sensor_gp_cov}
      \\
      &\mathbf{K}_{i,k} = \mathbf{P}^{-}_{g_{j},k}  \mathbf{H}_{i}^{T} \mathbf{R}_{i,k}^{-1},
      \label{eqn_sensor_gp_gain}
    \\
    & \hat{\mathbf{x}}^{-}_{g_{j},k} = \hat{\mathbf{x}}_{q,k-1} + \sum\limits_{i=1}^{I_j} {\mathbf{K}_{i,k} ( \mathbf{y}_{i,k} - \mathbf{H}_{i} \hat{\mathbf{x}}_{q,k-1})}.
    \label{eqn_sensor_gp_estimate}
\end{align}
\STATE \textbf{Step 2}: Each sensor group $j$ encrypts its estimate $\enc{\hat{\mathbf{x}}_{{g_{j}},k}^{-}}$ and sends its encrypted state to the aggregator.
\STATE \textbf{Step 3}: Diffusion update at the aggregator:
\begin{align}
    & \mathbf{P}^{-}_{a,k} = \left( \sum\limits_{j=1}^{J} {(\mathbf{P}^{-}_{g_{j},k})^{-1}} \right)^{-1},
    \label{eqn_sensor_gp_agg_cov}
    \\
    & \enc{\hat{\mathbf{x}}^{-}_{a,k}} = \mathbf{P}^{-}_{a,k} \sum\limits_{j=1}^{J} {(\mathbf{P}^{-}_{g_{j},k})^{-1} \enc{\hat{\mathbf{x}}^{-}_{g_{j},k}}}.
    \label{eqn_sensor_gp_agg_estimate}
\end{align}
\STATE \textbf{Step 4}: Time update at the aggregator:
\begin{align}
    & \enc{\hat{\mathbf{x}}_{a,k}} = \mathbf{F}  \enc{\hat{\mathbf{x}}_{a,k}^{-}},
     \label{eq_diff_est_tUpdt}
        \\
    & \mathbf{P}_{a,k} = \mathbf{F} \mathbf{P}_{a,k}^{-} \mathbf{F}^{T} + \mathbf{Q}_{k}.
    \label{eq_diff_cov_tUpdt}
\end{align}
\STATE \textbf{Step 5}: The aggregator sends the encrypted state $\enc{\hat{x}_{a,k}}$, $\mathbf{P}_{a,k}$ to the query node which decrypts and sends the results to the sensor groups.
\end{algorithmic}
\end{algorithm}
In the following theorem, we summarize the privacy guarantees of Protocol \ref{protocol_2} against coalitions predefined in Definitions
\ref{def_sensor_caol}, \ref{def_cloud_caol}, and \ref{def_query_caol}:
\begin{theorem}
\label{th_diff_setup}
Protocol \ref{protocol_2} solves Problem \ref{prob2_stmnt} while preserving computational privacy against
\begin{itemize}
    \item Sensor coalitions,
    \item Cloud coalitions,
    \item Query coalitions if $d_r>1$, where $d_r$ is the number of non-colluding groups.
\end{itemize}
\end{theorem}

Proof of Theorem \ref{th_diff_setup} is detailed in Appendix \ref{app2}.
\section{Privacy Discussion} \label{sec_privacy_discussion}
This paper demonstrates that Protocol \ref{protocol_1} solves Problem \ref{prob1_stmnt} with reasonable privacy guarantees summarized in Theorem \ref{th_dist_setup}, which illustrates that the protocol preserves privacy against all sensor and cloud coalitions. In the case of query coalitions, it is necessary to constantly make sure that the number of non-colluding sensors $m_r$ is greater than the ratio of the state size to the measurement size $n/p$ to preserve privacy. Besides, we can suggest a minor change by keeping the measurement matrix $\mathbf{H}_{i}$ or covariance $\mathbf{R}_{i,k}$ private to the sensors and aggregator to overcome this information leakage.

Similarly, Protocol \ref{protocol_2} solves Problem \ref{prob2_stmnt} while obtaining satisfying privacy guarantees as summarized in Theorem \ref{th_diff_setup}, which states that privacy is preserved against all coalitions unless the query node succeeds in colluding with all but one of the groups in an attempt to reveal the estimates of that group. Therefore, we must ensure that the number of non-colluding groups $d_r$ is always more than one. To overcome the information leakage in the case of the query coalitions, we propose a slight modification by keeping the process matrix $\mathbf{F}$ or the modeling noise covariance $\mathbf{Q}_{k}$ private to the aggregator. 

\vspace{-2mm}
\begin{remark}
There is an analogy between our protocols and two \textit{privacy-preserving set-based estimation} protocols proposed in \cite{sets_est_privacy} 
using \textit{Zonotopes}. Zonotope $\mathcal{Z}= \zono{c,G}$ is a centerally symmetric set representation, where $c$ is its center and $G$ is its generator matrix. Assuming that the modeling and measurement noises are unknown but bounded by zonotopes: $\mathbf{n}_k \in \mathcal{Z}_{\mathbf{Q},k} = \zono{0, \mathbf{Q}^z_k}$ and $\mathbf{v}_{i,k} \in \mathcal{Z}_{\mathbf{R},k} = \zono{0, \mathbf{R}^z_k}$ respectively, and having $\hat{\mathbf{x}}_k \equiv \hat{\mathbf{c}}_k$, $\mathbf{P}_k \equiv \hat{\mathbf{G}}_k \hat{\mathbf{G}}_k^T$, $\mathbf{R}_{i,k} \equiv \mathbf{R}^z_{i,k} {\mathbf{R}^z}_{i,k}^T$ \cite{sets_diff},
we found that our first protocol achieves privacy guarantees similar to the privacy guarantees of the set-based estimation protocol among distributed sensors \cite[Theorem 6.2]{sets_est_privacy}, while privacy guarantees of our second protocol are similar to their peers of the set-based estimation among sensor groups \cite[Theorem 7.2]{sets_est_privacy}.
\end{remark}

\section{Evaluation} \label{sec_evaluation}

\begin{figure}
\centering
\subfloat[Protocol \ref{protocol_1}] {\includegraphics[width=0.49\linewidth]{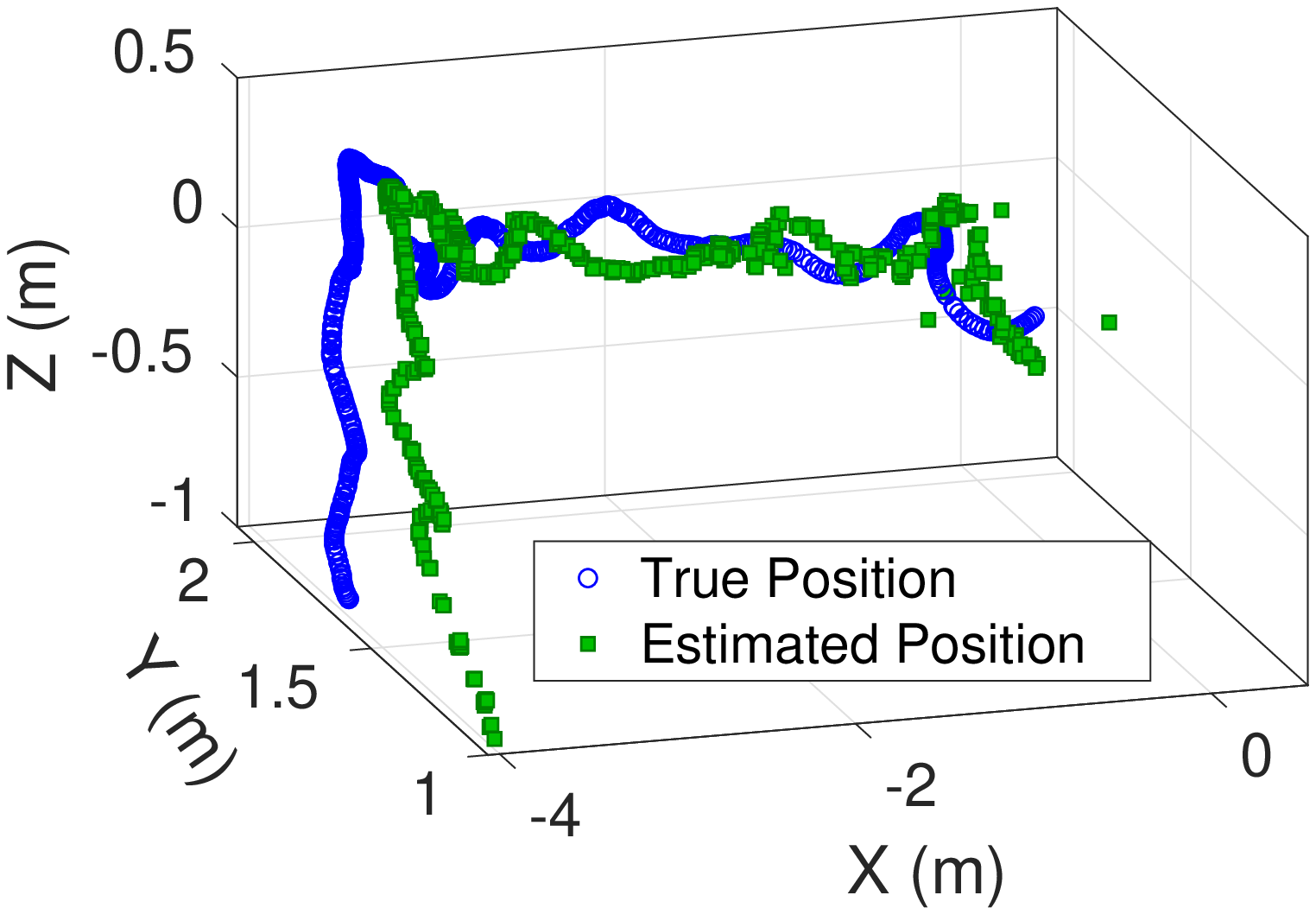}}
\hfill
\subfloat[Protocol \ref{protocol_2}]{ \includegraphics[width=0.49\linewidth]{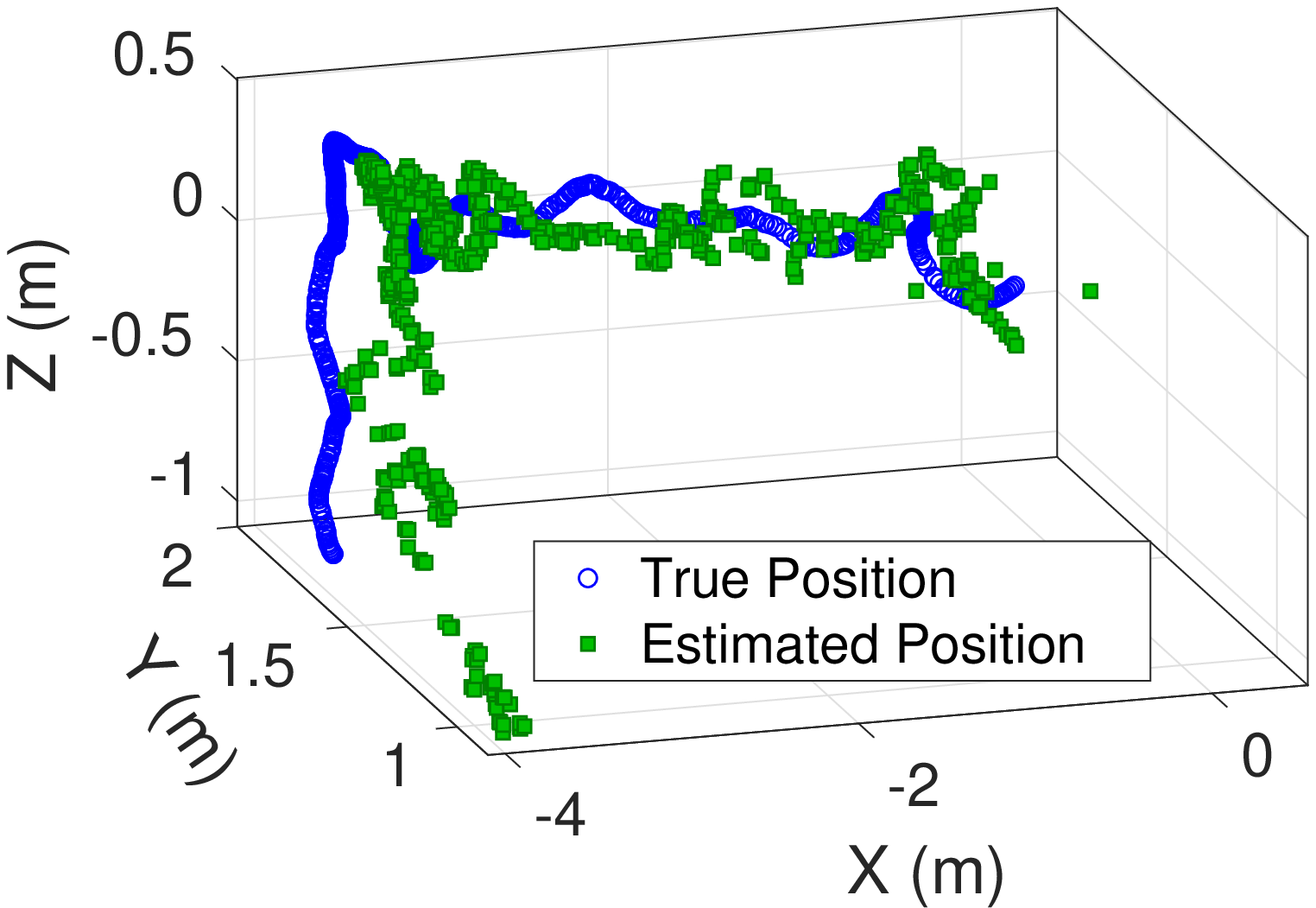}}
\vspace{-2mm}
\caption{True and estimated positions of the two proposed protocols.}
\label{fig_est}
\vspace{-3mm}
\end{figure}

\begin{figure}
\centering
\subfloat[Protocol \ref{protocol_1}] {\includegraphics[width=0.49\linewidth]{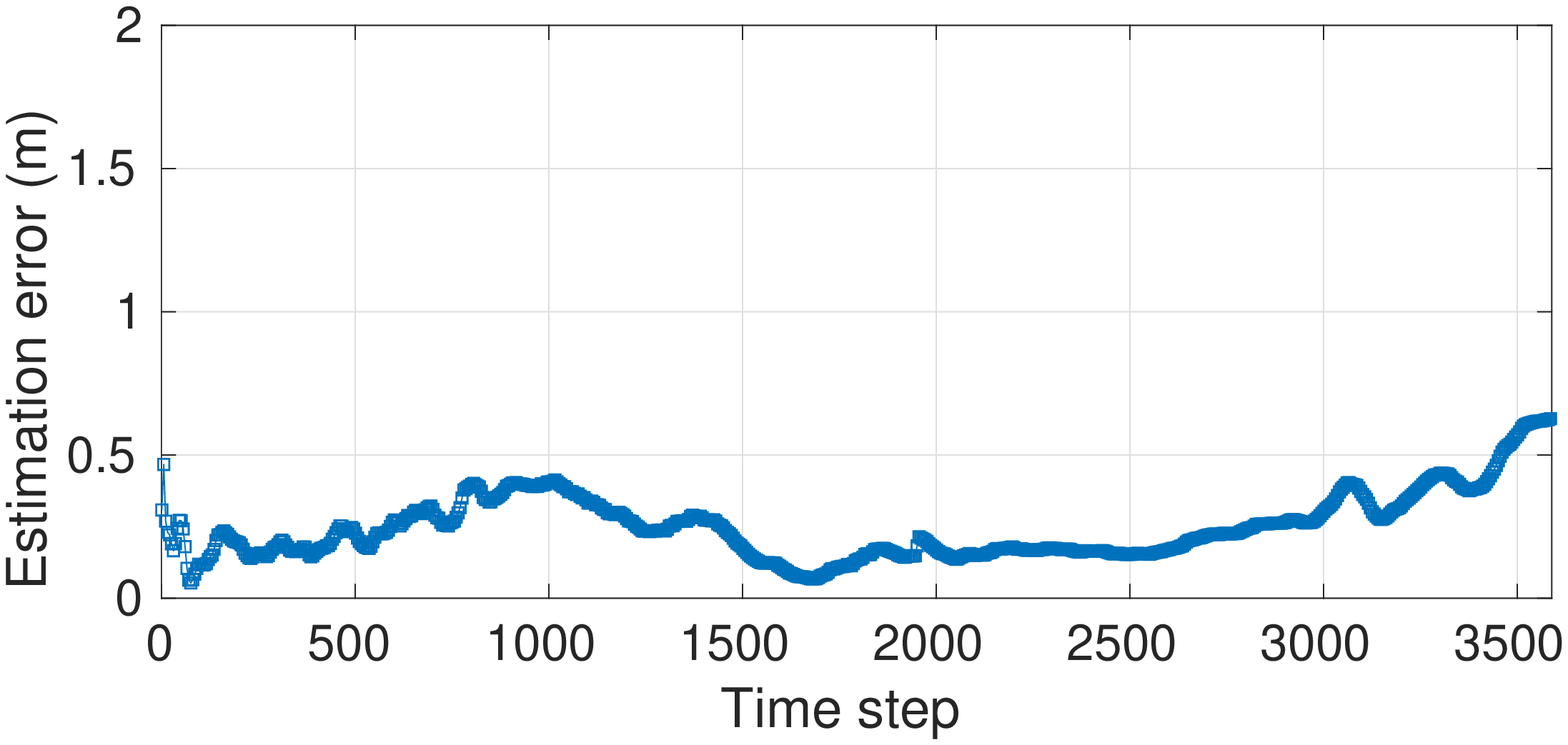}}
\hfill
\subfloat[Protocol \ref{protocol_2}]{ \includegraphics[width=0.49\linewidth]{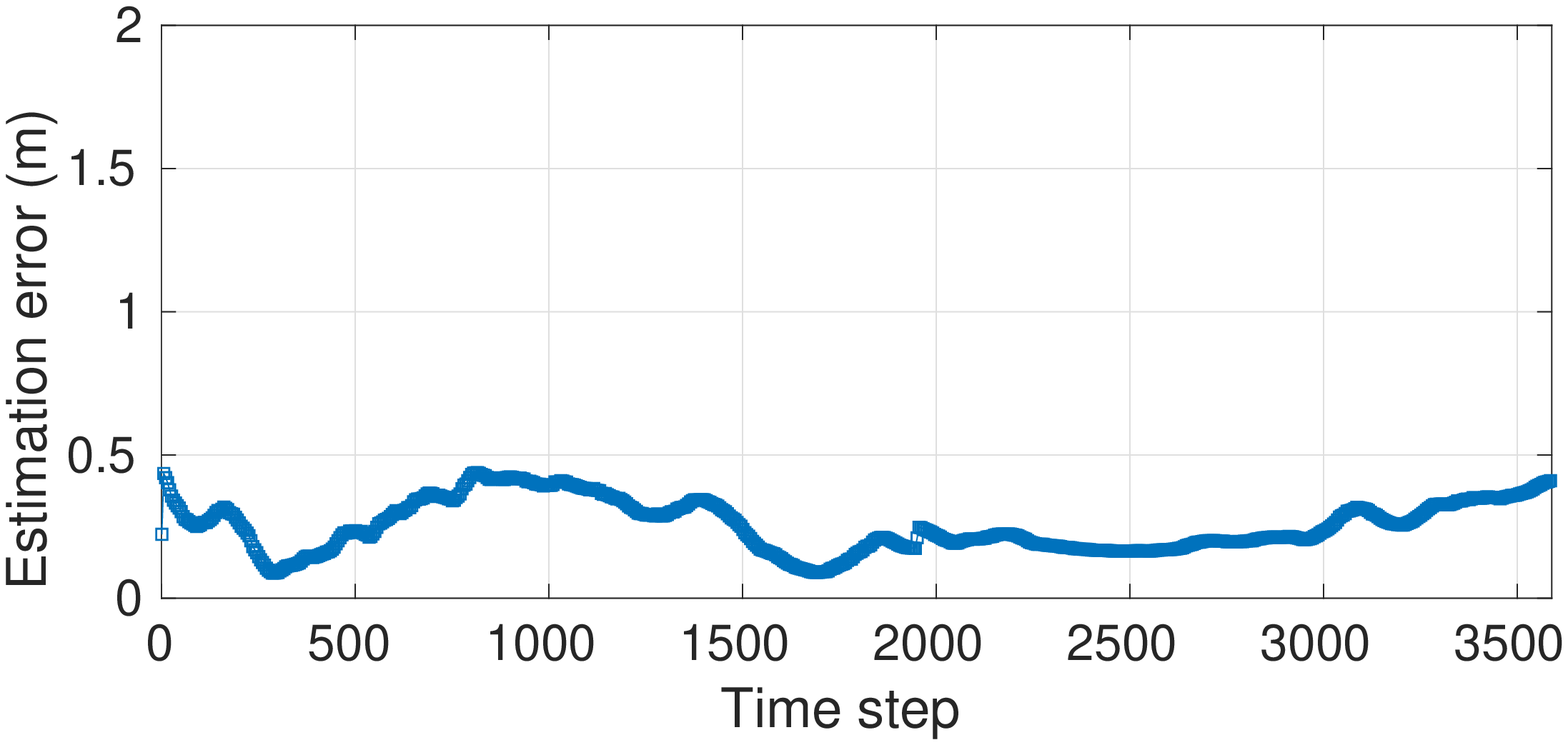}}
\vspace{-2mm}
\caption{Estimation error of the two proposed protocols.}
\label{fig_esterr}
\vspace{-1mm}
\end{figure}

To evaluate the proposed protocols, we use measurements collected from the real testbed used in \cite{ferraz2017node,alanwar2020event}, which includes a motion capture system that provides 3D rigid body position measurements. We apply our protocols to estimate the location of a quadrotor while preserving our previously mentioned privacy goals. Fig. \ref{fig_est} shows the true and estimated values of the 3D positions for each protocol.
The estimation error of the two protocols is presented in Fig. \ref{fig_esterr}. 
A comparison between the average execution time of each party is presented in Table \ref{table_exectime}. 
The cloud time in Protocol \ref{protocol_2} is shorter and depends on the diffusion method, and the encryption time is longer in sensor groups based on the sizes of state and measurement.





\begin{table}[t]
\caption{Execution Time in \SI{}{\milli\second}.}
\label{table_exectime}
\vspace{-2mm}
\centering
\normalsize
\begin{tabular}{c  c c c}
\toprule
   & \multicolumn{3}{c}{Entities}\\
  \cmidrule(lr){2-4} 
  & Sensor/Sensor group & Aggregator & Query\\
\midrule
Protocol \ref{protocol_1}  & 7.75   & 4.6  & 1.77 \\
Protocol \ref{protocol_2}  & 23.67  & 2.31  & 1.77 \\
\bottomrule
\end{tabular}
\vspace{-2mm}
\end{table}

\section{Conclusions} \label{sec_conclusions}

This paper presented two privacy-preserving estimation protocols for both a typical sensor setup and another setup in which trustworthy sensors are grouped into sensor groups. We proved the privacy guarantees of each protocol using the computational indistinguishability concept. We demonstrated the possibility of encrypting only sensitive information rather than all while maintaining an acceptable level of privacy. Finally, we evaluated our protocols using actual data from a physical testbed and verified that they offer satisfying results while guaranteeing privacy. Our protocols have several practical applications, which we leave as future work.



\bibliographystyle{ieeetr}
\bibliography{main} 

\appendix

\vspace{-1mm}
\subsection{Theorem \ref{th_dist_setup} Proof}
\label{app1}
The proof is along similar lines of \cite{sets_est_privacy}. 
To prove that the privacy is preserved, 
 we need to show that the coalitions views and simulators are computationally indistinguishable and that the coalition inputs and outputs do not leak extra private information according to Definition \ref{def_semihonest}. The quantities denoted by $\widetilde{()}$ are those obtained by the simulator and they differ from the quantities of the views but follow the same distribution. "Coins" are random numbers used for the encryption process and key generation. $\enc{\Gamma_X}$ for coalition $X$ represents any information that is transferred between other parties over an encrypted channel that may use double encryption with different keys from the homomorphic encryption keys. 
\vspace{-0.5mm}
 \begin{proof}
 \label{privacy_proof_1}
 In the following proof, we investigate the privacy of Protocol \ref{protocol_1} against the following three coalitions:
 \subsubsection{Coalition of sensors $s$} \label{sec_sensor_coalition}
If we consider a set of sensors $s=\{s_{1},\dots,s_{t}\}$ participating in the coalition $s$, then $V^{\Pi}_{s}$ represents the coalition view that can be defined as a combination of every sensor view and given by
\begin{align}
    V^{\Pi}_{s} &= \big(V^{\Pi}_{s_{1}},\dots,V^{\Pi}_{s_{t}} \big) \nonumber \\
    &= \big( \mathbf{H}_{s,k},\mathbf{y}_{s,k},\mathbf{R}_{s,k}, \enc{\mathbf{y}_{s,k}}, \textit{pk}, \text{coins}_{s}, \enc{\Gamma_s}
    \big), 
    \label{eq_view_sensor_coalition}
\end{align}
and the coalition simulator $S_s$ can be obtained by generating 
$\widetilde{\enc{\Gamma_s}}, \widetilde{\enc{y_{s,k}}}$ and  $\widetilde{\text{coins}}_{s}$, i.e.,
\begin{align}
    S_{s} =& \big(\mathbf{H}_{s,k},\mathbf{y}_{s,k},\mathbf{R}_{s,k}, \widetilde{\enc{\mathbf{y}_{s,k}}}, \textit{pk}, \widetilde{\text{coins}}_{s}, \widetilde{\enc{\Gamma_s}}
    \big),
\end{align}
where $\widetilde{\text{coins}}_{s}$ are generated according to the same distribution of $\text{coins}_{s}$ and are independent from other parameters, where the same is true for $\widetilde{\enc{\mathbf{y}_{s,k}}}$ and $\enc{\mathbf{y}_{s,k}}$ as well as $\widetilde{\enc{\Gamma_s}}$ and $\enc{\Gamma_s}$. 
Therefore, we find that $S_{s} \stackrel{c}{\equiv} V^{\Pi}_{s}$.
Furthermore, each measurement information is independent of all others, which makes the coalition unable to infer new information about the measurements of non-colluding sensors. For this coalition, we consider a single step in our proof since the information in each iteration differs from the other iterations. 

For the next two coalitions, we will prove that each coalition view after $K\in\mathbb{N}^+$ iterations of the protocol is computationally indistinguishable from the view of a simulator that runs the same $K$ iterations.
\subsubsection{Coalition of sensors and aggregator $sa$}
\label{sec_sensor_aggr_coalition}

Considering $V^{\Pi}_{a}$ is the view of the aggregator, the view of a coalition consisting of a set of sensors $s=\{s_{1},\dots,s_{t}\}$ and the aggregator can be indicated by $V^{\Pi}_{sa}$ where
\begin{align}
    V^{\Pi}_{sa} =& \big( V^{\Pi}_{s},V^{\Pi}_{a} \big) = \big( V^{\Pi,K}_{s},V^{\Pi,K}_{a} \big),
    \label{eq_view_sensor_aggr_pr1}
\end{align}
 where $V^{\Pi,K}_{s}$ and $V^{\Pi,K}_{a}$ are the views of the coalition of sensors and the aggregator respectively after executing $K$ iterations, and 
\begin{align}
\label{eq_add_instant_views}
 V^{\Pi,k+1}_{s}&=( V^{\Pi,k}_{s},I_{s}^{k+1}), \quad \nonumber \\
 V^{\Pi,k+1}_{a}&=( V^{\Pi,k}_{a},I_{a}^{k+1}), \quad k=0, 1,\dots, K-1,
\end{align}
where $I_{s}^k$ and $I_{a}^k$ are the new data added at the $k$-th iteration for the sensors coalition $V^{\Pi,0}_{s}=I_{s}^0$ and the aggregator $V^{\Pi,0}_{a}=I_{a}^0$. The view of the aggregator includes the sensors encrypted measurements and the initial estimates from the query node. The measurements at the $k$-th iteration from sensors that are not part of the coalition are marked by subscript $r$ (i.e., $\mathbf{H}_{r,k},\enc{\mathbf{y}_{r,k}},\mathbf{R}_{r,k}$, $k=0, 1,\dots, K-1$). Then, assuming that $\mathbf{H}_{r,k},\mathbf{R}_{r,k}$ are public, $I_{s}^k$ and $I_{a}^k$ are 
\begin{align}
    I_{s}^k = \big( &\mathbf{H}_{s,k},\mathbf{y}_{s,k},\mathbf{R}_{s,k},\textit{pk},\enc{\mathbf{y}_{s,k}},\text{coins}_{s}, \enc{\Gamma_s}
    \big),
    \label{eq_Instant_view_sensors} \\
    I_{a}^k =\big( &\mathbf{H}_{s,k},\enc{\mathbf{y}_{s,k}},\mathbf{R}_{s,k}, \mathbf{H}_{r,k},\enc{\mathbf{y}_{r,k}},\mathbf{R}_{r,k}, \enc{\hat{\mathbf{x}}_{q,0}}, \nonumber \\ 
    &{\mathbf{P}}_{q,0}, \enc{\hat{\mathbf{x}}_{a,k}}, {\mathbf{P}}_{a,k}, \mathbf{F}, \mathbf{Q_k}, \textit{pk}, \text{coins}_{a} \big), 
    \label{eq_Instant_view_aggr}
\end{align}
%
%
where $\enc{\hat{\mathbf{x}}_{q,0}}, \mathbf{P}_{q,0}$ is the initial estimates from the query node and $\enc{\hat{\mathbf{x}}_{a,k}},\mathbf{P}_{a,k}$ is the $k$ estimates on the aggregator side. 
The view of the coalition $V^{\Pi}_{sa}$ is constructed from \eqref{eq_view_sensor_aggr_pr1}-\eqref{eq_Instant_view_sensors}. The simulator of the coalition can be denoted by $S_{sa} =S_{sa}^{K}$, where $S_{sa}^{K}$ is the simulator after executing $K$ iterations. The simulator $S_{sa}$ can be formed using
 \begin{align}
S_{sa}^{k+1}=(S^{k}_{sa},I^{S,k+1}_{sa} ),\quad k=0, 1,\dots, K-1,
 \end{align}
 where $I^{S,k+1}$ is the simulator portion generated at iteration $k+1$, that is given by
\begin{align}
I^{S,k}_{sa}=\big( & \mathbf{H}_{s,k},\widetilde{\enc{\mathbf{y}_{s,k}}},\mathbf{R}_{s,k}, \mathbf{H}_{r,k},\widetilde{\enc{\mathbf{y}_{r,k}}},\mathbf{R}_{r,k}, \widetilde{\enc{\hat{\mathbf{x}}_{q,0}}}, \mathbf{p}_{q,0}, \nonumber \\ & \widetilde{\enc{\hat{\mathbf{x}}_{a,k}}}, \mathbf{P}_{a,k}, \mathbf{F}, \mathbf{Q_k}, \mathbf{y}_{s,k},\textit{pk}, \widetilde{\text{coins}}_{sa},\widetilde{\enc{\Gamma_s}}
\big),
\end{align}
where its terms are generated or computed as follows:
\begin{itemize}
    \item Generate $\widetilde{\enc{\mathbf{y}_{s,k}}}$, $\widetilde{\enc{\mathbf{y}_{r,k}}}$, $\widetilde{\enc{\hat{\mathbf{x}}_{q,0}}}$, $\widetilde{\enc{\hat{\mathbf{x}}_{a,k}}}$ and $\widetilde{\enc{\Gamma_s}},$ according to the same distribution of $\enc{\mathbf{y}_{s,k}}$, $\enc{\mathbf{y}_{r,k}}$, $\enc{\hat{\mathbf{x}}_{q,0}}$, $\enc{\hat{\mathbf{x}}_{a,k}}$ and $\enc{\Gamma_s}$, respectively.
    \item Compute $\mathbf{P}_{a,k}$ according to \eqref{eqn_enc_kalman3}.
    \item Suppose both coins are combined as $\text{coins}_{sa} = (\text{coins}_{a},\text{coins}_{s})$, then generate $\widetilde{\text{coins}}_{sa}$ according to the same distribution.
\end{itemize}
%
Then, all the $\widetilde{\enc{}}$ and $\enc{}$ values are indistinguishable and all other variables in $I^{S,k+1}_{sa}$ are either public or attainable through the protocol steps. Hence, After all iteration steps, we end up with a simulator that achieves $S_{sa} \stackrel{c}{\equiv} V^{\Pi}_{sa}.$
%
%
Now we need to ensure the coalition cannot infer 
further private information. The coalition's target is 
to find the private measurements of the non-colluding sensors $\mathbf{y}_{r,k}$. The relation between $\enc{\mathbf{y}_{s,k}}$ and $\enc{\mathbf{y}_{r,k}}$ can be derived from \eqref{eqn_enc_kalman5} as
\begin{align}
 \sum\limits_{i\in\mathcal{N}_{r}} \mathbf{K}_{i,k} \enc{\mathbf{y}_{i,k}} = & \sum\limits_{i\in\mathcal{N}_{I}} ( \mathbf{K}_{i,k} \mathbf{H}_{i,k} - 1) \enc{\hat{\mathbf{x}}_{a,k-1}} \oplus \enc{\hat{\mathbf{x}}_{a,k}} \nonumber \\
  & \ominus \underbrace{\sum\limits_{i\in \mathcal{N}/r} \mathbf{K}_{i,k} \enc{\mathbf{y}_{i,k}}}_{\text{known to the coalition in plaintext}}, \label{eq_unknowns_sensor_aggr_col}
\end{align}
where $\mathcal{N}_{r}$ is the remaining sensors set
. Since the coalition does not have the private key and the query node sends the initial encrypted estimate $\enc{\hat{\mathbf{x}}_{a,0}}$,As a result, we have a system that is undermined in \eqref{eq_unknowns_sensor_aggr_col}.
\subsubsection{Coalition of sensors and query node $sq$} \label{sec_sensors_query_col}
Let $V^{\Pi}_{q}$ be the view of the query node, then, the view of a coalition consisting of a set of sensors 
and the query is $V^{\Pi}_{sq}$ that defined by
\begin{align}
V^{\Pi}_{sq} =& \big( V^{\Pi}_{s},V^{\Pi}_{q} \big) = \big( V^{\Pi,K}_{s},V^{\Pi,K}_{q} \big),
\end{align}
where 
\begin{align}
\label{eq_add_Instant_views_sensors_query}
V^{\Pi,k+1}_{s}&=( V^{\Pi,k}_{s},I_{s}^{k+1}), \quad \nonumber\\ 
V^{\Pi,k+1}_{q}&=( V^{\Pi,k}_{q},I_{q}^{k+1}),\quad k=0, 1,\dots, K-1,
\end{align}
where $I_{s}^k$ is given in \eqref{eq_Instant_view_sensors}, and $I_{q}^k$ are the new data added from the $k$-th iteration for the query node with $ V^{\Pi,0}_{q}=I_{q}^0$ such that
\begin{align}
\label{eq_Instant_view_Query}
I_{q}^k=\big(\hat{\mathbf{x}}_{q,0},\mathbf{P}_{q,0},\enc{\hat{\mathbf{x}}_{a,k}},\mathbf{P}_{a,k},\hat{\mathbf{x}}_{a,k},\textit{pk},\textit{sk}, \text{coins}_{q}, \enc{\Gamma_{sq}}\big).
\end{align}
The coalition view $V^{\Pi}_{sq}$ can be formed using \eqref{eq_Instant_view_sensors}, \eqref{eq_add_Instant_views_sensors_query} and \eqref{eq_Instant_view_Query}, where the simulator can be constructed 
as

\begin{align}
    S_{sq}^k = \big( &\mathbf{H}_{s,k},\mathbf{y}_{s,k},\mathbf{R}_{s,k},\hat{\mathbf{x}}_{q,0},{P}_{q,0},\hat{\mathbf{x}}_{a,k},\mathbf{P}_{a,k},\nonumber \\
    &\widetilde{\enc{\hat{\mathbf{x}}_{a,k}}},\widetilde{\text{coins}}_{sq},\widetilde{\enc{\Gamma_{sq}}}, \textit{pk},\textit{sk}, S_{sq}^{k-1} \big),
\end{align}
where $\left(\widetilde{\enc{\hat{\mathbf{x}}_{a,k}}}, \widetilde{\text{coins}}_{sq}, \widetilde{\enc{\Gamma_{sq}}}\right)$ are generated according to the same distribution of $\left(\enc{\hat{\mathbf{x}}_{a,k}}, \text{coins}_{sq}, \enc{\Gamma_{sq}}\right)$ and are independent from other parameters. Therefore, $S_{sq}^k \stackrel{c}{\equiv} (I_{s}^k,I_{q}^k),$ which proves that 
  $S_{sq} \stackrel{c}{\equiv} V^{\Pi}_{sq}$.
To complete our proof and similar to \cite{sets_est_privacy}, it is important to examine whether the coalition can reveal the private information of the non-colluding $m_r$ sensors. Since the query has the Paillier private key \textit{sk}, we can rewrite \eqref{eq_unknowns_sensor_aggr_col} after decryption
 \begin{align}\label{eq_privacy}
\textbf{K}_{r,k} \textbf{Y}_{r,k}= \mathbf{z}_{s,k},
 \end{align}
where
\begin{align}
\label{eq_knowns_sensor_query_col}
&\mathbf{z}_{s,k}=\sum\limits_{i\in\mathcal{N}_{I}} ( \mathbf{K}_{i,k} \mathbf{H}_{i,k}) \hat{\mathbf{x}}_{a,k-1} + \hat{\mathbf{x}}_{a,k} - \sum\limits_{i\in \mathcal{N}/r } \mathbf{K}_{i,k} \mathbf{y}_{i,k}, 
\end{align}
\begin{align}
&\textbf{K}_{r,k}=[\mathbf{K}_{i_1,k},\mathbf{K}_{i_2,k},\dots,\mathbf{K}_{{i_{m_r}},k}]\in\mathbb{R}^{n\times pm_r},\nonumber \\
&\textbf{Y}_{r,k}=[\mathbf{y}_{i_1,k}^T,\mathbf{y}_{i_2,k}^T,\dots,\mathbf{y}_{{i_{m_r}},k}^T]^T\in\mathbb{R}^{ pm_r}, \nonumber
\end{align}
where $\mathbf{z}_{s,k}$ is known to the coalition since each $\textbf{K}_{i,k}$ can be calculated using \eqref{eqn_enc_kalman4}. We need to examine whether there is no unique retrieval for $Y_{r,k}$ to preserve its privacy, which means that \eqref{eq_privacy} has multiple solutions. 
According to \cite[Theorem 6.4]{schott2016matrix}, $\tilde{Y}_{r,k}$ is a solution of \eqref{eq_privacy} for any $X_r \in\mathbb{R}^{pI_r}$ with
 \begin{align}\label{eq_privacy1_1}
\tilde Y_{r,k}=\textbf{K}_{r,k}^\textbf{--}\mathbf{z}_{s,k}+(I_{pm_r}-\textbf{K}_{r,k}^\textbf{--}\textbf{K}_{r,k})X_r,
 \end{align}
where $\textbf{K}_{r,k}^\textbf{--}$ is any generalized inverse of $\textbf{K}_{r,k}$. 
In order to ensure that \eqref{eq_privacy1_1} has multiple solutions, we need to keep $(I_{pm_r}-\textbf{K}_{r,k}^\textbf{--}\textbf{K}_{r,k})\neq 0$. 
Having $\textbf{K}_{r,k}^\textbf{--}\textbf{K}_{r,k} \neq I_{pm_r}$ implies that $rank(\textbf{K}_{r,k} )<pm_r$, which is always true while $n<pm_r$ according to \cite[Theorem 5.23, Theorem 2.6]{schott2016matrix}
. Then, under the condition $pm_r>n$, 
system \eqref{eq_privacy1_1} has infinity solutions,
which preserves the privacy of $Y_{r,k}$.
\end{proof}

\subsection{Theorem \ref{th_diff_setup} Proof}
\label{app2}
%

 \begin{proof}
  \label{privacy_proof_2}
Along the same line of the previous proof, we consider the view and simulator for one step (i.e., $k$-th step).
The proof for $K\in\mathbb{N}^+$ steps is the same as previous proof. We prove again the privacy against the following three coalitions:
 \subsubsection{Coalition of sensor groups $g$}
%
Let $V^{\Pi}_{g}$ denote the view of a coalition includes a set of sensor groups $g=\{g_{1},\dots,g_{t}\}$ and defined by
\begin{align}
    V^{\Pi}_{g} = \big( &V^{\Pi}_{g_{1}},\dots,V^{\Pi}_{g_{t}} \big) \\
    = \big( &\mathcal{H}_{g,k}, \mathcal{Y}_{g,k}, \mathcal{R}_{g,k}, \mathbf{P}_{q,k-1}, \hat{\mathbf{x}}_{q,k-1}, \mathbf{P}^{-}_{g,k}, \nonumber \\
    &\hat{\mathbf{x}}^{-}_{g,k}, \textit{pk}, \enc{\hat{\mathbf{x}}^{-}_{g,k}}, coins_g, \enc{\Gamma_g}
    \big),
\end{align}
where the subscript $g$ denotes items owned by the coalition. Each sensor group submit only its encrypted prior estimate to the aggregator. Hence, the simulator $S_g$ is defined by
\begin{align}
        S_{g} = \big( &\mathcal{H}_{g,k}, \mathcal{Y}_{g,k}, \mathcal{R}_{g,k}, \mathbf{P}_{q,k-1}, \hat{\mathbf{x}}_{q,k-1}, \mathbf{P}^{-}_{g,k}, \nonumber \\
        &\hat{\mathbf{x}}^{-}_{g,k}, \textit{pk},  \widetilde{\enc{\hat{\mathbf{x}}^{-}_{g,k}}}, \widetilde{coin}_g, \widetilde{\enc{\Gamma_g}} \big),
\end{align}
and $\widetilde{\enc{\hat{\mathbf{x}}^{-}_{g,k}}}$,  $\widetilde{\text{coins}}_{g}$ and $\widetilde{\enc{\Gamma_g}}$ are generated with the same distribution of $\enc{\hat{\mathbf{x}}^{-}_{g,k}}, coins_g, \enc{\Gamma_g}$ and are independent from other parameters. Therefore, we find that $ S_{g} \stackrel{c}{\equiv} V^{\Pi}_{g}$. Also, the coalition cannot infer any further information about other estimates of the non-colluding groups.
%
\subsubsection{Coalition of sensor groups and the aggregator $ga$} 
The coalition view is defined by
\begin{align}
    V^{\Pi}_{ga} =& \big( V^{\Pi}_{g},V^{\Pi}_{a} \big),
\end{align}
where
\begin{align}
    V^{\Pi}_{a} = \big( & \enc{\hat{\mathbf{x}}^{-}_{g,k}}, \mathbf{P}^{-}_{g,k}, \enc{\hat{\mathbf{x}}^{-}_{r,k}}, \mathbf{P}^{-}_{r,k}, \enc{\hat{\mathbf{x}}_{a,k}}, \mathbf{P}_{a,k}, \mathbf{F},
    \nonumber \\ 
    & \mathbf{Q_k}, \text{coins}_a, \textit{pk} \big).
\end{align}
The simulator $S_{ga}$ can be constructed by calculating $\mathbf{P}^{-}_{g,k}$ and $\mathbf{P}^{-}_{r,k}$ using \eqref{eqn_sensor_gp_cov}, $\mathbf{P}^{-}_{a,k}$ using \eqref{eqn_sensor_gp_agg_cov}, and generating $\widetilde{\enc{\hat{\mathbf{x}}^{-}_{g,k}}}, \widetilde{\enc{\hat{\mathbf{x}}^{-}_{r,k}}}, \widetilde{\enc{\hat{\mathbf{x}}_{a,k}}}$, and $\widetilde{\enc{\Gamma_g}}$ as before.
\begin{align}
    S_{ga} = \big( &\mathcal{H}_{r,k}, \mathcal{R}_{r,k}, \mathcal{H}_{g,k}, \mathcal{Y}_{g,k}, \mathcal{R}_{g,k},\mathbf{P}_{q,k-1}, \hat{\mathbf{x}}_{q,k-1}, \nonumber \\
    &\mathbf{P}^{-}_{g,k}, \hat{\mathbf{x}}^{-}_{g,k}, {P}^{-}_{r,k}, \widetilde{\enc{\hat{\mathbf{x}}^{-}_{r,k}}}, \widetilde{\enc{\hat{\mathbf{x}}^{-}_{g,k}}},\widetilde{\enc{\hat{\mathbf{x}}_{a,k}}}, \mathbf{P}_{a,k}, \nonumber \\
     &F, Q_k, \textit{pk}, \widetilde{\text{coins}}_{ga},  \widetilde{\enc{\Gamma_g}} \big).
\end{align}
Thus, we find that $ S_{ga} \stackrel{c}{\equiv} V^{\Pi}_{ga}$. To ensure that the coalition cannot find the remaining groups estimates, we use \eqref{eqn_sensor_gp_agg_estimate} to describe the relation between group estimates as
 \begin{align}
 \sum\limits_{j\in\mathcal{N}_{r}}{(\mathbf{P}^{-}_{g_{j},k})^{-1} \enc{\hat{\mathbf{x}}^{-}_{g_{j},k}}} = &(\mathbf{P}^{-}_{a,k})^{-1} \enc{\hat{\mathbf{x}}^{-}_{a,k}} \nonumber
 \\ &\ominus \underbrace{\sum\limits_{j\in\mathcal{N}/r} {(\mathbf{P}^{-}_{g_{j},k})^{-1} \enc{\hat{\mathbf{x}}^{-}_{g_{j},k}}}}_{\text{known to the coalition in plaintext}}, \label{eq_unknowns_sensorgps_aggr_col}
\end{align}

where ${N}_{r}$ is the set of non-colluding groups with size $d_r$. 
However, since the coalition does not know the private key, the privacy of the remaining groups' estimates can be guaranteed.

\subsubsection{Coalition of sensor groups and the query node $gq$}
The coalition view $V^{\Pi}_{gq}$ is defined as $V^{\Pi}_{gq} = \big( V^{\Pi}_{g},V^{\Pi}_{q} \big)$ where 
\begin{align}
    V^{\Pi}_{q} =& \big( \enc{\hat{\mathbf{x}}_{a,k}},\hat{x}_{a,k}, \mathbf{P}_{a,k}, 
 \textit{pk},\textit{sk}, \text{coins}_q, \enc{\Gamma_q} \big).
\end{align}
And the simulator $S_{gq}$ can be constructed as before 
\begin{align}
S_{gq} = \big(&\mathcal{H}_{g,k}, \mathcal{Y}_{g,k}, \mathcal{R}_{g,k}, \mathbf{P}^{-}_{g,k}, \hat{\mathbf{x}}^{-}_{g,k}, \nonumber \\  
&\widetilde{\enc{\hat{\mathbf{x}}_{a,k}}},\hat{\mathbf{x}}_{a,k}, \mathbf{P}_{a,k}, \textit{pk}, \textit{sk}, \widetilde{\text{coins}}_{gq}, \widetilde{\enc{\Gamma_q}} \big).
\end{align}
Thus we conclude that $S_{gq} \stackrel{c}{\equiv} V^{\Pi}_{gq}$. Similar to \cite{sets_est_privacy} and to investigate the privacy of the non-colluding groups' estimates, we use \eqref{eq_unknowns_sensorgps_aggr_col} after decryption as the query has the private key \textit{sk}.
 \begin{align}\label{eq_privacy_protocol_2}
\textbf{P}_{r,k} \textbf{X}_{r,k}= \mathbf{z}_{g,k},
 \end{align}
with
\begin{align}
\mathbf{z}_{g,k}&=(\mathbf{P}^{-}_{a,k})^{-1} \hat{\mathbf{x}}^{-}_{a,k} - 
    \sum\limits_{j\in\mathcal{N}/r} {(\mathbf{P}^{-}_{g_{j},k})^{-1} \hat{\mathbf{x}}^{-}_{g_{j},k}}, \\
\textbf{P}_{r,k} &=[ (\mathbf{P}^{-}_{j_1,k})^{-1}, (\mathbf{P}^{-}_{j_2,k})^{-1},\dots, (\mathbf{P}^{-}_{j_{d_r},k})^{-1}]\in\mathbb{R}^{n\times nd_r}, \nonumber \\
\textbf{X}_{r,k}&=[\mathbf{x}_{j_1,k}^T, \mathbf{x}_{j_2,k}^T,\dots,\mathbf{x}_{j_{d_r},k}^T]^T\in\mathbb{R}^{ nd_r}, \nonumber
\end{align}
where $z_{g,k}$ is known to the coalition as $\hat{\mathbf{x}}^{-}_{a,k}$ and $\mathbf{P}^{-}_{a,k}$ can be calculated using \eqref{eq_diff_est_tUpdt} and \eqref{eq_diff_cov_tUpdt} assuming that $\mathbf{F}$ and $\mathbf{Q}_{k}$ are public and $\mathbf{F}$ is invertible. If $\mathbf{F}$ isn't invertible, then the privacy of the remaining groups will be guaranteed against all coalitions.
Similarly to proof \ref{privacy_proof_1} and according to \cite[Theorem 6.4]{schott2016matrix}, $\tilde{X}_{r,k}$ is a solution of \eqref{eq_privacy_protocol_2} for any $X_r \in\mathbb{R}^{nd_r}$ where
 \begin{align}\label{eq_privacy2}
\tilde X_{r,k}=\textbf{P}_{r,k}^\textbf{--} \mathbf{z}_{g,k} + ({I}_{nd_r}-\textbf{P}_{r,k}^\textbf{--} \textbf{P}_{r,k} ) X_r,
 \end{align}
with $\textbf{P}_{r,k}^\textbf{--}$ is any generalized inverse of $\textbf{P}_{r,k}$.
We aim to find conditions at which $I_{nd_r}-\textbf{P}_{r,k}^\textbf{--}\textbf{P}_{r,k}\neq0$ to ensure privacy. 
Having $\textbf{P}_{r,k}^\textbf{--}\textbf{P}_{r,k}\neq I_{nd_r}$ implies that $rank(\textbf{P}_{r,k})<nd_r$, which is always true while 
$n<nd_r$ according to \cite[Theorem 5.23, Theorem 2.6]{schott2016matrix}
. Thus, under the condition $d_r>1$, the privacy of $X_{r,k}$ is guaranteed.
%
\end{proof}

\end{document}